A causal role of sensory cortices
in behavioral benefits of 'learning by doing'

*[†]Brian Mathias[1,2], [†]Leona Sureth[2], Gesa Hartwigsen[3],

Manuela Macedonia[2,4], Katja M. Mayer[5], & Katharina von Kriegstein[1,2]

[1]Chair of Cognitive and Clinical Neuroscience, Faculty of Psychology, Technical University
Dresden, Dresden, Germany
[2]Research Group Neural Mechanisms of Human Communication, Max Planck Institute for
Human Cognitive and Brain Sciences, Leipzig, Germany
[3]Lise Meitner Research Group Cognition and Plasticity, Max Planck Institute for Human
Cognitive and Brain Sciences, Leipzig, Germany
[4]Linz Center of Mechatronics, Johannes Kepler University Linz, Linz, Austria
[5]Department for Psychology, University of Münster, Münster, Germany

[†]Joint first authors

*Corresponding author:

Brian Mathias

Technical University Dresden

Chair of Cognitive and Clinical Neuroscience

Faculty of Psychology

Bamberger Str. 7

01187 Dresden

Germany

Email: brian.mathias@tu-dresden.de

**Abstract**

Despite a rise in the use of "learning by doing" pedagogical methods in praxis, little is known as to how these methods improve learning outcomes. Here we show that visual association cortex causally contributes to performance benefits of a learning by doing method. This finding derives from transcranial magnetic stimulation (TMS) and a gesture-enriched foreign language (L2) vocabulary learning paradigm performed by 22 young adults. Inhibitory TMS of visual motion cortex reduced learning outcomes for abstract and concrete gesture-enriched words in comparison to sham stimulation. There were no TMS effects on words learned with pictures. These results adjudicate between opposing predictions of two neuroscientific learning theories: While reactivation-based theories predict no functional role of visual motion cortex in vocabulary learning outcomes, the current study supports the predictive coding theory view that specialized sensory cortices precipitate sensorimotor-based learning benefits.

Foreign language (L2) vocabulary learning by adults is effortful and time-consuming. Words must be relearned often to build up robust memory representations. L2 vocabulary learning typically relies on unisensory material such as written word lists or audio recordings (Choo, Lin, & Pandian, 2012). More recent learning-by-doing-based approaches contrast with these techniques. Though initially viewed as unconventional, principles of learning by doing have shifted from the periphery of educational science toward its center over the past few decades. Such strategies make use of visual and somatosensory information as well as motor information. We will therefore in the following refer to learning by doing strategies as *sensorimotor-enriched* teaching. Sensorimotor-enriched teaching methods boost test performance in science, engineering, mathematics, and L2 learning compared to other teaching methods (Freeman et al., 2014; for a review see Macedonia, 2014). Underlying human brain mechanisms that support beneficial effects of sensorimotor-enriched teaching methods are currently unknown.

A recent behavioral and functional magnetic resonance imaging (fMRI) study investigated sensorimotor-enriched L2 vocabulary learning using four enriched teaching methods and non-enriched control conditions (Mayer, Yildiz, Macedonia, & von Kriegstein, 2015). In the first enrichment condition, learners heard L2 words and their native language (L1) translations while performing iconic gestures that were modelled by an actress in video clips (gesture *performance* enrichment). In the second enrichment condition they heard a different set of L2 words while viewing videos without performing the gestures (gesture *viewing* enrichment). Gesture performance enrichment but not gesture viewing enrichment enhanced post-training learning outcomes compared to non-enriched L2 learning. Interestingly, the third enrichment condition, which entailed viewing iconic pictures during learning (picture *viewing* enrichment), also enhanced post-training outcomes compared to non-enriched L2 learning. However, the fourth enrichment condition, in which pictures were traced by hand (picture *performance* enrichment) did not benefit post-training performance compared to non-enriched L2 learning. Gesture-performance-enriched and picture-viewing-enriched learning outcomes did not differ from each other immediately after the 5-day training protocol. Gesture performance enrichment, however, led to more favorable learning outcomes than picture viewing enrichment over the long-term (6 months post-training).

Mayer and colleagues' (2015) fMRI results demonstrated that L2 words presented in the auditory modality following training could be decoded in specialized visual association cortices on the basis of how the words were learned: The biological motion area of the superior temporal sulcus (bmSTS), a region implicated in visual perception of biological movements (Grossman et al., 2000), was engaged by words that had been learned under both types of gesture enrichment (i.e., gesture performance and gesture viewing). The lateral occipital complex (LOC) was instead engaged following picture-enriched learning (picture performance and picture viewing). These neuroimaging results are consistent with studies showing that the presence of complementary sensory information during learning elicits reactivation of specialized sensory brain regions at test (Danker & Anderson, 2010; Lahav, Saltzman, & Schlaug, 2007).

The reactivation of visual brain regions elicited by stimuli presented in the auditory modality may be viewed as epiphenomenal, a view taken by reactivation theories of multisensory learning (Fuster, 2009; Nyberg, Habib, McIntosh, & Tulving, 2000; Wheeler, Petersen, & Buckner, 2000). For example, if the taste of a cake triggered the recall of a visual memory, the recollection might reactivate visual areas; reactivation theories, however, assume that those visual responses do not aid in making the sensory experience of the cake's taste more precise. Within a reactivation-based framework, benefits of sensorimotor enrichment on learning could be relegated to arousal-based effects that are not dependent on representations subserved by sensory brain regions.

Reactivation theories can be seen as a critical counterpart to the notion that sensorimotor networks formed during real-world experience support perception (for reviews see Barsalou, 2010, von Kriegstein 2012, and Matusz, Wallace, & Murray, 2017). According to this alternate view, sensory brain responses to previously-learned items directly benefit recognition processes by increasing recognition speed and accuracy. The predictive coding theory of multisensory learning (Mayer et al., 2015; von Kriegstein & Giraud 2006; for review see von Kriegstein, 2012) takes this approach by proposing that sensory and motor cortices build up sensorimotor (e.g., visuomotor) forward models during perception that predict or simulate missing input, which functionally benefit behavioral learning outcomes. If that were the case, teaching techniques could be optimized to target specific sensory structures that underlie task performance. The first aim of the current study was thus to adjudicate between tenets of reactivation and predictive coding theories of multisensory learning. The second aim of the study was to evaluate the role of multisensory brain responses not only immediately following learning, but also over extended post-training durations. This aim was based on the finding that gesture-performance enrichment has shown to benefit learning over longer time scales than the initially equally effective viewing of pictures (Mayer et al., 2015),

Though functional neuroimaging has contributed much to our understanding of interactions between information arising from distinct sensory modalities (for review see James & Stevenson, 2012), neuroimaging techniques are limited to demonstrations of correlational rather than causal effects (Ramsey et al., 2010). Conversely, transcranial magnetic stimulation (TMS) is a key method for making causal claims regarding the role of specialized sensory and motor areas in behavioral benefits of sensorimotor-enriched learning. TMS can be used to test whether brain regions that are engaged during sensorimotor-enriched learning *causally* contribute to its beneficial behavioral effects. During TMS, small magnetic fields are applied non-invasively to the scalp, targeting a specific brain area. The magnetic fields induce electrical currents in the underlying brain tissue, transiently interfering with processing in a specific area. If the stimulated region is causally relevant for an ongoing task, then an observable behavioral effect, usually an increase in response latencies, can be induced (Sack et al., 2007). TMS cannot be substituted by behavioral, fMRI, or other correlational neuroscience measurements.

## Methods

### Overview and Hypotheses

The study was constructed in close analogy to the design by Mayer et al (2015). Adult learners were trained on 90 novel L2 words and their L1 translations over 4 consecutive days (**Fig. 1a**). L2 vocabulary (concrete and abstract nouns, supplementary **Table S1**) was learned in two conditions. In a gesture-performance-enriched learning condition, individuals viewed and performed gestures while L2 words were presented auditorily. In a picture-viewing-enriched learning condition, individuals viewed pictures while L2 words were presented auditorily (**Fig. 1b**). Gestures and pictures were congruent with word meanings. We selected the gesture performance enrichment and picture viewing enrichment conditions for two main reasons. First, of the four enrichment conditions previously tested in adults (Mayer et al., 2015), only these two conditions benefitted post-training L2 translation compared to auditory-only learning. Second, benefits of gesture performance enrichment and picture viewing enrichment were associated with responses in two different visual cortical areas, i.e., the bmSTS for gesture performance enrichment and the LOC for picture viewing enrichment (Mayer et al., 2015). For succinctness, we hereafter refer to the gesture performance enrichment condition as the "gesture enrichment" learning condition, and the picture viewing enrichment condition as the "picture enrichment" learning condition. We used TMS to target the bmSTS, as the bmSTS is more easily accessible for TMS than the LOC (Siebner & Ziemann, 2007). TMS was applied to the bilateral bmSTS while participants translated auditorily-presented L2 words into L1 at two time points: 5 days and 5 months following the start of L2 training. Participants did not perform gestures or view pictures during the TMS task. A within-participants control condition was included in each TMS session by applying both effective and sham TMS to the bilateral bmSTS (**Fig. 1c**).

Classification accuracy of neural responses within the bmSTS previously correlated with performance in a multiple choice translation task (Mayer et al., 2015). In this task, participants selected the correct L1 translation of an auditorily-presented L2 word from a list of options presented on a screen. This task was also used for the present TMS design and we refer to it as the "multiple choice task" (**Fig. 1d**). In addition, we included an exploratory recall task in the present TMS design in which participants pressed a button as soon as the L1 translation came to their mind when hearing each L2 word at the start of each trial (not shown in **Fig. 1d**, see procedure and supplementary results for more details).

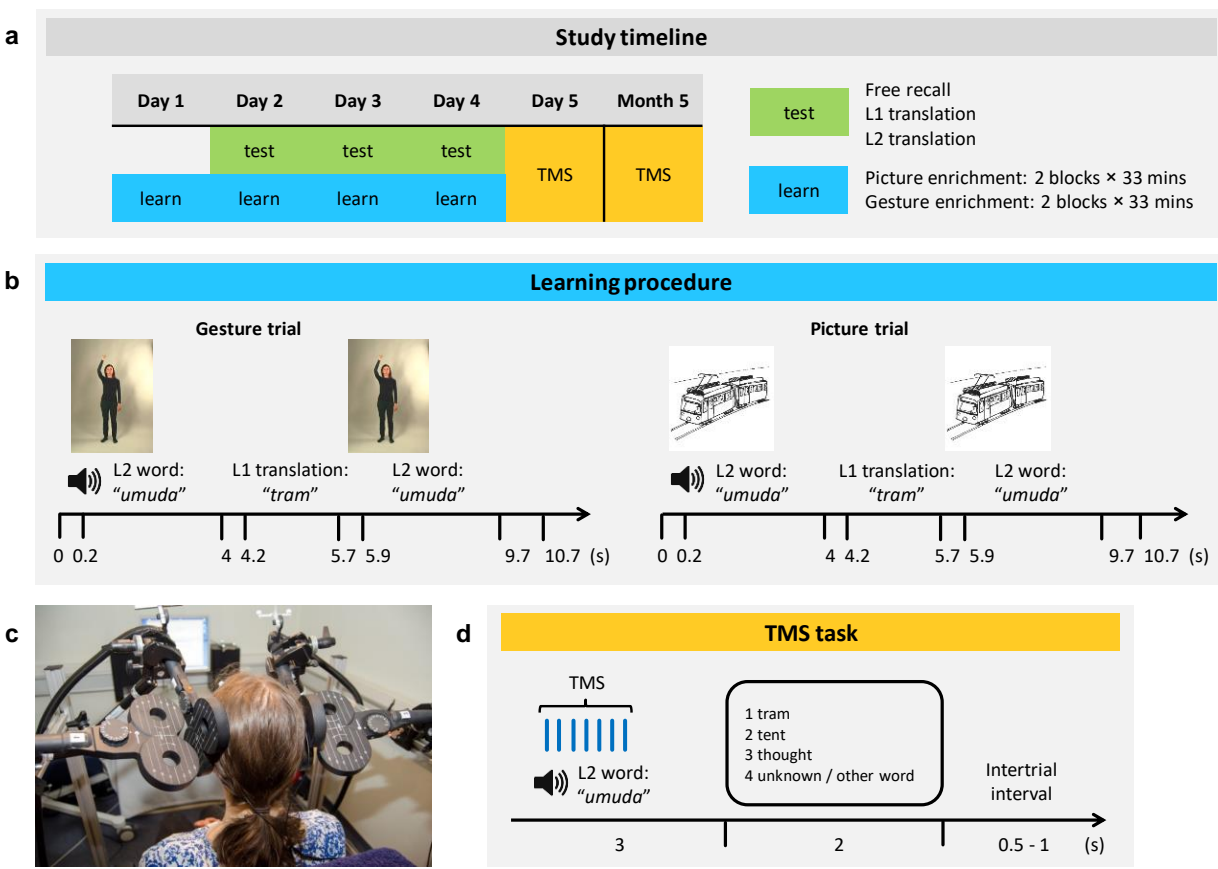

**Figure 1. Experimental procedure and design. a**, Participants learned foreign language (L2) vocabulary over four consecutive days ('learn') in groups to emulate a classroom setting. Free recall and translation tests ('test') were administered on days 2 through 4. Transcranial magnetic stimulation (TMS) sessions occurred during day 5 and month 5. **b**, In both gesture and picture learning conditions, participants heard an L2 word, followed by the translation in their native language (L1) and a repetition of the L2 word. Videos of iconic gestures and pictures

accompanied L2 words in gesture and picture trials, respectively. Participants performed the gesture along with the video during its repetition. **c**, During both TMS sessions (day 5 and month 5), two TMS coils targeted the bilateral biological motion superior temporal sulcus (bmSTS) using stereotactic neuronavigation based on individual structural brain scans. Two additional coils generated ineffective placebo stimulation (i.e., sham TMS) and were positioned on top of the bmSTS coils at an angle of 90°. **d**, During each TMS session, participants heard the L2 words that they had learned during the 4-day training and then selected the L1 translation by button press from a list of options presented on a screen (multiple choice task). L1 words were presented in German. Trains of seven TMS pulses at 10 Hz were delivered 50 ms following each L2 word onset. Trials with effective and sham TMS alternated in blocks. Note that participants additionally pressed a button as soon as the L1 translation came to their mind following each auditorily-presented L2 word (recall task; not shown, see methods and supplementary results for more details).

We tested three main hypotheses. First, according to the predictive coding theory of multisensory learning (von Kriegstein, 2012), the application of inhibitory stimulation to the bmSTS should slow down the translation of an auditorily-presented L2 word in comparison to sham stimulation if the word has been learned with biological motion, as was the case in our gesture-enrichment condition. There should be no such effect if the word has been learned with pictures. Thus we expected an interaction between learning condition (gesture, picture) and stimulation condition (effective, sham). The interaction should be driven by a simple main effect of stimulation condition on the translation of gesture-enriched words. In contrast, reactivation learning theories (Fuster, 2009; Nyberg et al., 2000; Wheeler et al., 2000), which assume that reactivated areas do not play a functional role in recognition, would predict no differential effects of bmSTS stimulation in contrast to sham bmSTS stimulation on the translation of auditorily-presented L2 words (i.e., there would be no interaction between learning and TMS conditions).

Our second hypothesis was that bmSTS integrity would support the auditory translation of gesture-performance-enriched words at the later time point (5 months post-learning) even more than the earlier time point (5 days post-learning). This hypothesis was based on the finding that gesture-performance enrichment is particularly powerful for learning outcome on time-scales of several months post learning in comparison to picture-viewing enrichment (Mayer et al., 2015). In our design, this hypothesis could be tested by examining the interaction between learning condition, stimulation type, and time point variables: We expected greater effects of bmSTS stimulation compared to sham stimulation on translation response times for gesture-enriched words at the later time point than at the earlier time point, and no effects of bmSTS stimulation compared to sham stimulation for picture-enriched words at either time point.

Our third hypothesis was that bmSTS stimulation would yield similar effects on the translation of both concrete and abstract words (see supplementary introduction on influences of the conceptual perceptibility of L2 word referents). This prediction was based on previous results showing that sensorimotor enrichment can benefit the learning of both word types (Macedonia, 2014; Mayer et al., 2015).

Response time was used as the dependent variable for testing our three hypotheses, because response time is the standard measure for TMS tasks: TMS typically influences response times rather than accuracy (Ashbridge, Walsh, & Cowey, 1997; Hartwigsen et al., 2017; Pascual-Leone et al., 1996; Sack et al., 2007).

In addition to testing our three main hypotheses, the design allowed us to test the reliability of the previously-reported finding that benefits of gesture performance enrichment exceed those of picture viewing enrichment over the long-term (Mayer et al., 2015). To this end, we examined accuracy outcomes in the multiple choice task, and predicted that overall accuracy would be greater for gesture-enriched words compared to picture-enriched words 5 months post-learning.

## Participants

Twenty-two right-handed native German speakers (15 females; *M* age = 26.6 years, *SD* = 4.7 years, range 18-35) completed the study. The sample size was based on two previous experiments (*n* = 22 per experiment) that estimated beneficial effects of gesture and picture enrichment on foreign language (L2) learning outcomes (Mayer et al., 2015, Experiments 1 and 2).

None of the participants reported a history of neurological or psychiatric disorders, head injury, or any contraindications for TMS. All participants reported normal hearing and normal or corrected-to-normal vision. None of the participants were raised in bilingual households. Of 32 total participants who registered for the study, one participant experienced an adverse reaction to TMS (convulsive syncope) and did not complete the testing. Syncope is the most common adverse reaction to TMS; the exact prevalence is unknown (Rossi, Hallett, Rossini, & Pascual-Leone, 2009). Another participant completed the training sessions but did not start the initial TMS session for medical reasons. Four participants were excluded because they were unable to return for additional testing sessions that occurred 5 months following the initial testing sessions due to time constraints, and 4 other participants were excluded due to poor localization of right or left bmSTS coordinates in individual participant space.

Written informed consent was obtained from all participants prior to the study. Participants were informed that the goal of the study was to test the effectiveness of different vocabulary learning strategies in adulthood but they were naïve to the specific hypotheses. All participants were evaluated by a medical doctor prior to the study in order to be approved for TMS and magnetic resonance imaging (MRI). The study was approved by the ethics committee of the University of Leipzig.

**Stimuli**

Stimuli consisted of 90 pseudowords (see supplementary material, **Table S1**). The pseudowords were derived from an artificial foreign language corpus referred to as 'Vimmi', developed by Macedonia and colleagues (Macedonia & Knösche, 2011) and intended for use in experiments on L2 learning. The corpus was created in order to control for participants' prior knowledge of foreign languages and for differences between words (e.g., length, frequency) in natural languages. Vimmi words conform to rules of Italian phonotactics (words sound like Italian but do not exist in the Italian language). All Vimmi words used in the current study were composed of three syllables consisting of vowels and consonants.

The 90 Vimmi words and 90 German translations used in the current study were previously evaluated by Mayer and colleagues (2015). Half of the 90 words were concrete nouns and the other half were abstract nouns. Lengths of concrete and abstract German words did not differ (concrete $M$ = 2.40 syllables, $SD$ = 0.84 syllables; abstract: $M$ = 2.69 syllables, $SD$ = 0.90 syllables). Frequency of the concrete and abstract words in written German did not differ (concrete frequency score: $M$ = 11.00, $SD$ = 1.18, range 9 to 13; abstract frequency score: $M$ = 10.96, $SD$ = 0.98, range: 9 to 13) (http://wortschatz.uni-leipzig.de/de).

**Videos, pictures, and audio files.** For each of the 90 Vimmi words, a 4 s color video was created using a Canon Legria HF S10 camcorder (Canon Inc., Tokyo, Japan). In each video, an actress performed a gesture that conveyed a word meaning. The actress was always positioned in the center of the video recording. She performed the gestures using head movements, movements of one or both arms or legs, fingers, or combinations of these body parts and maintained a neutral facial expression throughout each video. The word "bottle", for example, was represented by the actress miming drinking from an imaginary bottle, and the word "good deed" was represented by the actress miming laying a donation in the imaginary hat of a homeless individual. The actress began and ended each gesture by standing motionless with her arms at her sides. Large gestures (e.g., steps or jumps) were restricted to a 1 m radius around the body's starting position. Gestures used to convey the meanings of abstract words were agreed upon by 3 independent parties (Mayer et al., 2015).

A black-and-white line drawing was created by a professional illustrator for each of the 90 Vimmi words. Pictures conveyed word meanings by portraying humans, objects, or scenes. Pictures illustrating concrete nouns were mostly drawings of single objects, and pictures illustrating abstract nouns were often scenes. The complexity of the illustrations for concrete and abstract words was not matched, since similar differences are expected in natural teaching settings.

The same actress that carried out the gestures in the videos spoke the Vimmi and the German words in audio recordings. Words were recorded using a Rode NT55 microphone (Rode Microphones, Silverwater, Australia) in a sound-damped chamber. The actress is an Italian native speaker and recorded the Vimmi words with an Italian accent to highlight the foreign language aspect of the stimuli for German-speaking participants. Vimmi audio stimuli ranged from 654-850 ms in length ($M$ = 819.7 ms, $SD$ = 47.3 ms). For more details on the video, picture and audio files used in the current study, see Mayer and colleagues (2015). Sample stimuli can be found at http://kriegstein.cbs.mpg.de/mayer_etal_stimuli/.

**Experimental Design**

The study utilized a 2 × 2 × 2 × 2 repeated-measures design. Within-participant independent factors were learning enrichment condition (gesture, picture), TMS condition (effective stimulation, sham stimulation), testing time point (5 days, 5 months), and L2 vocabulary type (concrete, abstract). Participants received both effective and sham TMS within the same session at each time point. The order in which effective and sham TMS conditions were administered within each session was counterbalanced across participants within each time point and between time points.

**Procedure**

**L2 vocabulary learning.** Participants learned L2 words in two conditions. In the gesture learning condition, individuals viewed and performed gestures while L2 words were presented auditorily. In the picture learning condition, individuals viewed pictures while L2 words were presented auditorily. Each day of learning comprised four 33-minute learning blocks. Blocks alternated between gesture and picture enrichment conditions. Each of the 45 Vimmi words included in a single learning block was repeated 4 times per block, yielding a total of 180 randomly-ordered trials per block. Participants took breaks of 10 minutes between blocks.

During breaks, participants conversed with each other and consumed snacks and drinks that were provided. Enrichment condition orders were counterbalanced across participants and learning days.

Participants were instructed prior to the start of learning that the goal was to learn as many Vimmi words as possible over the 4 days of training. Participants received no further instruction during the training except to be informed about which learning condition would occur next (i.e., gesture or picture enrichment). Since the L2 vocabulary learning took place in groups of up to 4 individuals, training sessions occurred in a seminar room with a projector and a sound system. Audio recordings were played via speakers located on each side of the screen. The volume of the playback was adjusted so that all participants could comfortably hear the words.

The assignment of the 90 stimuli to learning enrichment conditions was counterbalanced such that half of the participants learned one set of 45 words in the gesture learning condition and the other set of 45 words in the picture learning condition. The other half of the participants received the reverse assignment of stimuli to gesture and picture conditions. This manipulation ensured that each Vimmi word was equally represented in both the gesture enrichment condition and the picture enrichment control condition.

In each gesture enrichment trial (**Fig. 1b**), participants first heard an L2 word accompanied by a video of an actress performing a gesture that conveyed the meaning of the word (shown for 4 s). They then heard the native language (L1) translation paired with a blank screen. Finally, the L2 word was presented a second time, again accompanied by the same video of the actress performing the gesture. Participants were asked to enact the gesture along with the actress during the second showing of each video. They were free to perform the gestures mirror-inverted or they could use their right arm when the actress in the video used her right arm, for example; they were asked to use only one of the two strategies throughout the learning period. In each picture enrichment trial (**Fig. 1b**), participants first heard an L2 word accompanied by a picture that conveyed the meaning of the word (shown for 4 s). They then heard the L1 translation paired with a blank screen. Finally, the same L2 word was presented a second time, again accompanied by the same picture. A motor task was not included in the picture enrichment condition as the enrichment of picture viewing with motor information (e.g., tracing an outline of presented pictures) has been shown to be less beneficial for learning than simply viewing the pictures without performing a motor task (Mayer et al., 2015). We therefore did not combine picture enrichment with motor performance in the current study. Participants stood during all learning blocks. Standing locations during the training were counterbalanced over the 4 learning days.

On days 2, 3, and 4 of the L2 vocabulary learning, participants completed paper-and-pencil vocabulary tests prior to the training, shown in **Fig. 1a**. We included these tests in order to maintain the same L2 training procedure used by Mayer et al (2015). Participants completed free recall, L1 translation, and L2 translation tests on each day. More information on the vocabulary tests as well as the test results can be found in the supplementary material. The participant with the highest combined scores on the paper-and-pencil vocabulary tests across days 2, 3, and 4 was rewarded with an additional 21€ beyond the total study compensation of 211€. Participants were informed about the financial incentive on day 1 prior to the start of the learning blocks.

Prior to vocabulary learning on day 1, participants completed three psychological tests examining their concentration ability (Concentration test; Brickenkamp, 2002) (*M* score = 211.6, *SD* = 51.1), speech repetition ability (Nonword Repetition test; Korkman, Kirk, & Kemp, 1998) (*M* score = 98.2, *SD* = 8.8), and verbal working memory (Digit Span test; Neubauer & Horn, 2006) (*M* score = 18.7, *SD* = 3.7). None of the participants were outliers (3 *SD* above or below the group mean) with respect to their scores on any of the three tests, and all participants performed within the norms of the Concentration test for which norms were available. Participants also completed a questionnaire on their prior knowledge of foreign languages and language learning experience.

**TMS translation tasks.** Participants performed recall and multiple choice tasks (**Fig. 1d**) while undergoing effective and sham TMS in two TMS sessions (5 days and 5 months following the start of L2 vocabulary learning). The two tasks were performed in four 6-minute blocks, each containing 45 words that had been presented on days 1 to 4. Each of the 90 words learned during the learning days was presented twice per TMS session, for a total of 180 test trials per TMS session and task. Effective and sham stimulation alternated across blocks, with half of the participants receiving effective stimulation during the first block and the other half receiving sham stimulation during the first block. Stimuli were ordered randomly within effective and sham stimulation blocks.

Each trial began with the written instruction "Press the button as soon as you know the translation" presented for 1.5 s on a screen. This was followed by the auditorily-presented L2 word accompanied by a black screen. A train of seven TMS pulses at 10 Hz delivered to the bilateral bmSTS began 50 ms after the onset of each word. Participants responded as soon as they recalled the L1 translation of the L2 word by pressing a button with their right index finger (recall task, not shown in **Fig. 1d**). If they did not know the L1 translation, they did not respond. Three seconds following L2 word onsets, a screen with four response options appeared and participants were given up to 2 s to select the correct L1 translation (**Fig. 1d**). The fourth response option was always "Unknown / Other word"; participants were told to select this option if they did not know the L1 translation or thought that the correct translation was different from the three options presented. They responded by pressing one of four buttons on the response pad with their index, middle, ring, or little fingers (multiple choice task). Even if participants did not know the translation of the L2 word after hearing it, they were still able to select one of the four options presented. Responses were considered correct if participants pressed the correct button while the response screen was present. Participants were instructed to always respond as quickly and as accurately as possible. Each trial ended with a jittered inter-stimulus interval (0.5 to 1 s) paired with a black screen. Following the first TMS session, participants completed a questionnaire on strategies that they used to learn and remember the L2 words.

Several months following the first TMS session, participants were invited to participate in a second TMS session. The second session occurred approximately 18 weeks ($M = 18.0$ weeks, $SD = 1.4$ weeks) following the first session. Participants completed the same two tasks as during the first TMS session while again undergoing effective and sham stimulation. Following the second TMS session, participants completed a questionnaire on strategies they used to remember meanings of the L2 words during the second session.

Finally, participants returned to complete the pencil-and-paper vocabulary tests (free recall, L1 translation, and L2 translation) 2 to 6 days ($M = 4.1$ days, $SD = 1.3$ days) after their second TMS session. Participants had no knowledge of the additional TMS and behavioral sessions until they were contacted a few weeks prior to their 5-month target testing dates. This was done to avoid potential rehearsal of the vocabulary during the 5-month interval between testing time points.

**Transcranial Magnetic Stimulation**

**Neuronavigation.** Stereotactic neuronavigated TMS was performed using Localite software (Localite GmbH, Sankt Augustin, Germany). Neuronavigation based on structural neuroimaging data from individual participants allows precise positioning of TMS coils. T1-weighted MRI scans for each participant were obtained with a 3-Tesla MAGNETOM Prisma-fit (Siemens Healthcare, Erlangen, Germany) using a magnetization-prepared rapid gradient echo (MPRAGE) sequence in a sagittal orientation (repetition time = 2300 ms, echo time = 2.98 ms, inversion time = 900 ms, flip angle = 9°, voxel size = 1x1x1 mm).

Structural T1 brain scans used for TMS neuronavigation were obtained from all participants prior to the TMS sessions. During each TMS session, participants were co-registered to their T1 scans. The two stimulation coils used in the current study were placed over Localite-indicated entry points of the respective target sites on the scalp. Entry points were those coordinates on each participant's scalp that were the shortest distance to the target neural coordinates (right and left bmSTS). To stimulate the bmSTS bilaterally, a tangential coil orientation of 135° to the sagittal plane was applied with current flow within both stimulation coils reversed, resulting in a posterior to anterior (PA) current flow in the brain. A 135° coil orientation with a PA current flow is equivalent to a 45° coil orientation with an anterior to posterior (AP) current flow. Coils were secured in position using fixation arms (Manfrotto 244, Cassola, Italy).

Mean Montreal Neurological Institute (MNI) coordinates for bilateral bmSTS stimulation were derived from the functional MRI findings of Mayer and colleagues (2015): right bmSTS, *x, y, z* = 55, −41, 4; left bmSTS, *x, y, z* = −54, −41, −5. Mayer and colleagues (2015) found that participants translated auditorily-presented L2 words learned previously with gesture enrichment more accurately than L2 words learned without enrichment (auditory-only learning), referred to as a gesture enrichment benefit. Using multivariate pattern analysis (MVPA), they found that a classifier trained to discriminate BOLD responses to gesture-enriched and auditory-only words showed significant classification accuracy in the bmSTS. Classifier accuracy in the bmSTS positively correlated with the gesture enrichment benefit, suggesting a role of this area in improving learning outcomes following multisensory learning. In the current study, we stimulated the mean location across participants that demonstrated maximal classifier accuracy within the bmSTS. To ensure precise individual stimulation of target coordinates, mean MNI coordinates for the two target sites (right and left bmSTS) were transferred into individual subject space using SPM8 (Wellcome Trust Center for Neuroimaging, University College London, UK, http://www.fil.ion.ucl.ac.uk/spm/).

**TMS parameters.** Two MagPro X100 stimulators (MagVenture A/S, Farum, Denmark) and a total of four focal figure-of-eight coils (C-B60; outer diameter = 7.5 cm) were used for stimulation. Signal software version 1.59 (Cambridge Electronic Design Limited, Cambridge, UK) was used to control the TMS pulse sequence. Presentation software (Neurobehavioral Systems Inc., Berkeley, CA, USA) was used for stimulus delivery, response recording, and to trigger TMS pulses.

An EIZO 19" LCD monitor approximately 1 m in front of the seated participant displayed task-related text (white letters, font: Arial, font size: 32 pt; black background). Shure SE215 sound isolating in-ear headphones (Shure Europe, Eppingen, Germany) were used to deliver L2 word recordings during the TMS sessions. Sound volume was individually adjusted prior to beginning the TMS task.

During each TMS session, a within-participants control condition was included by applying not only effective TMS to the bilateral bmSTS but also sham TMS. Sham TMS coils for each hemisphere were positioned at a 90° angle over each stimulation coil, as shown in **Fig. 1c**, and therefore did not effectively stimulate the brain. Coil locations were monitored and adjusted for head movements during the TMS sessions. The repetitive TMS protocol used (a seven-pulse train of 10 Hz TMS) was in line with published TMS safety guidelines (Rossi et al., 2009).

Prior to the TMS translation task, each participant's individual stimulation intensity was determined by measuring their resting motor threshold (RMT). To measure RMT, we stimulated the hand region of the left primary motor cortex (M1) using single-pulse TMS, resulting in the conduction of motor-evoked potentials (MEPs) in the relaxed first dorsal interosseous muscle (FDI) of the right hand. The RMT was defined as the lowest stimulation intensity producing 5 MEPs out of 10 consecutive TMS pulses that exceeded a 50 mV peak-to-trough amplitude. A meta-analysis by Mayka and colleagues (2006) provided mean stereotactic coordinates of the left M1 ($x, y, z$ = −37, −21, 58 mm, MNI space), which were used as a starting point to locate the M1 FDI hotspot. The coil used to elicit MEPs was oriented at 45° to the sagittal plane, inducing a PA current flow in the brain.

Effective and sham TMS intensity during the L2 translation task was set to 90% of each participant's RMT. The same intensity was used for both TMS sessions for each participant ($M$ = 40.1% of maximum stimulator output, $SD$ = 5.6%).

**Data Analysis**

All participants who completed the study ($n$ = 22) were included in the analyses.

**Analysis of response times in the translation tasks.** Participants indicated that they recalled the L1 translation prior to the appearance of the four response options during fewer than half of all trials across the two TMS sessions ($M$ = 41.7% of trials, $SE$ = 4.5%), leaving an insufficient number of trials for analysis of the recall task. An exploratory analysis of these data can be found in the supplementary results. In contrast, in the multiple choice task, participants selected a translation from the multiple choice options presented on the screen during $M$ = 88.6% ($SE$ = 3.6%) of trials across the two TMS sessions. In the following we focus the analyses on the response times for the multiple choice task.

Response times in the multiple choice task were computed as the time interval from the appearance of the multiple choice options on the screen until the response. Trials in which participants did not respond following the appearance of the multiple choice options, selected the incorrect translation, or selected the fourth response options ("Unknown / Other word") were excluded from the response time analyses.

To test our first hypothesis (see overview and hypotheses subsection of the methods), we ran a two-way repeated measures analysis of variance (ANOVA) with the factors learning condition (gesture, picture) and stimulation type (effective, sham) on response times in the multiple choice task. To evaluate whether the observed patterns of response times were due to speed-accuracy tradeoffs, we correlated response times in the multiple choice translation task with accuracy (percent correct) for each learning condition, stimulation condition, and time point.

To test our second hypothesis (see overview and hypotheses subsection of the methods), we ran a three-way repeated measures ANOVA with factors learning condition (gesture, picture), stimulation type (effective, sham), and time point (day 5, month 5) on response times in the multiple choice task.

To test our third hypothesis (see overview and hypotheses subsection of the methods), we ran a four-way repeated measures ANOVA on response times in the multiple choice task with factors learning condition (gesture, picture), stimulation type (effective, sham), testing time point (day 5, month 5), and vocabulary type (concrete, abstract).

Pairwise comparisons for all analyses were conducted using two-tailed Tukey's honestly significant difference (HSD) post-hoc tests.

**Linear mixed effects modeling of response times in the multiple choice translation task.** To evaluate the robustness of the observed effects using an alternate analysis technique, we also tested our three hypotheses using a linear mixed effects modeling approach. We performed backwards model selection to select the model's random effects structure, beginning with a random intercept by subject, a random intercept by auditory stimulus, a random slope by subject for each of the four independent factors (stimulation type, learning condition, time point, and vocabulary type), and a random slope by stimulus for the stimulation type and time point factors. We removed random effects terms that accounted for the least variance one by one until the fitted mixed model was no longer singular, i.e., until variances of one or more linear combinations of random effects were no longer (close to) zero. The final model included three random effects terms: a random intercept by subject, a random intercept by stimulus, and a random slope by subject for the time point factor. Please see the supplementary material for more methodological details.

**Analysis of response accuracy.** Besides testing our main hypotheses, the data also allowed us to test the reliability of the previous finding that benefits of gesture performance enrichment on L2 translation exceeded those of picture viewing enrichment over the long-term (Mayer et al., 2015). We ran a four-way repeated measures ANOVA on accuracy in the multiple choice task, with the factors learning condition (gesture, picture), stimulation type (effective, sham), testing time point (day 5, month 5), and vocabulary type (concrete, abstract), and examined all interactions involving the learning and time point factors.

## Results

**Stimulation of the bmSTS slows the translation of gesture-enriched foreign vocabulary**

Our first and primary hypothesis was that a brain region specialized in the perception of biological motion, the bmSTS (Grossman et al., 2000), causally contributes to L2 translation following gesture-enriched L2 learning, but not picture-enriched L2 learning. We therefore first tested whether bmSTS stimulation modulated L2 translation, irrespective of testing time point. The results confirmed our hypothesis. A two-way ANOVA on response times in the multiple choice task revealed a stimulation type × learning condition interaction (F $_{1,21}$ = 11.82, $p$ = .002, two-tailed, $\eta_p^2$ = .36) (see **Table S3**, supplementary material, for the full set of ANOVA results). Tukey's HSD post-hoc tests revealed that response times for words that had been learned with gesture enrichment were significantly delayed when TMS was applied to the bmSTS compared to sham stimulation ($p$ = .005, Hedge's $g$ = .33). This was not the case for words learned with picture enrichment. This indicates that perturbation of a brain area related to biological motion slowed the translation of L2 words that had been learned with gestures, but not of L2 words learned with pictures (**Fig. 2**).

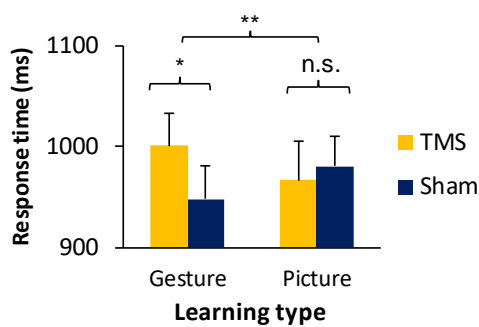

**Figure 2. Effects of bmSTS stimulation on speed of L2 translation.** Bilateral bmSTS stimulation slowed the translation of L2 vocabulary learned using gestures compared to sham stimulation in the multiple choice task. There was no such effect for L2 vocabulary learned using pictures. The mean of each condition across time points (5 days and 5 months following

the start of learning) is shown (*n* = 22 participants). Error bars represent one standard error of the mean. *$p$ < .05, **$p$ < .01.

In a control analysis, we tested whether differences in response times under effective stimulation compared to sham stimulation conditions could be due to tradeoffs between translation speed and accuracy. Response times for correct answers in the multiple choice translation task were correlated with accuracy (percent correct) for each learning condition, stimulation condition, and time point. If there were a speed-accuracy tradeoff, one would expect a positive correlation between response times and accuracy (i.e., the longer the response time, the greater the accuracy). Response times, however, did not correlate or correlated negatively with translation accuracy (**Table 1**). Thus, participants did not trade speed for accuracy.

|  | Day 5 | | Month 5 | |
|---|---|---|---|---|
|  | **TMS** | **Sham** | **TMS** | **Sham** |
|  | *r* (*p*) | *r* (*p*) | *r* (*p*) | *r* (*p*) |
| **Gesture** | -.84 (<.001)* | -.89 (<.001)* | -.63 (.002)* | -.34 (.12) |
| **Picture** | -.89 (<.001)* | -.95 (<.001)* | -.48 (.02) | -.46 (.03) |

**Table 1. Speed-accuracy relationships in L2 translation.** In most tests, slower response times correlated with lower translation accuracy, indicating that there was no speed-accuracy tradeoff. *df* = 20 for all correlations. *$p$ < .05, Bonferroni corrected.

**bmSTS supports auditory foreign vocabulary translation 5 months post-learning**

Our second hypothesis was that bmSTS integrity would support the auditory translation of gesture-enriched words at the later time point (5 months post-learning) even more than the earlier time point (5 days following the start of learning). In agreement with this hypothesis, a three-way ANOVA on response times for the multiple choice task yielded a significant three-way stimulation type × learning condition × time point interaction ($F_{1, 21}$ = 7.51, $p$ = .012, two-tailed, $\eta_p^2 = .26$) (see **Table S4**, supplementary material, for the full set of ANOVA results). Tukey's HSD post-hoc tests revealed a response benefit (faster responses) for words learned with gesture enrichment compared to words learned with picture enrichment under sham stimulation 5 months following learning ($p < .001$, Hedge's $g = .69$). The application of TMS to the bmSTS negated this benefit: Response times for gesture- and picture-enriched words did not significantly differ at month 5 under effective stimulation, and responses were significantly slower under effective stimulation compared to sham stimulation for words learned with gesture enrichment ($p = .001$, Hedge's $g = .61$). In sum, significant effects of bmSTS stimulation on translation were more prominent 5 months following the L2 training period compared to 5 days following the start of learning (**Fig. 3**).

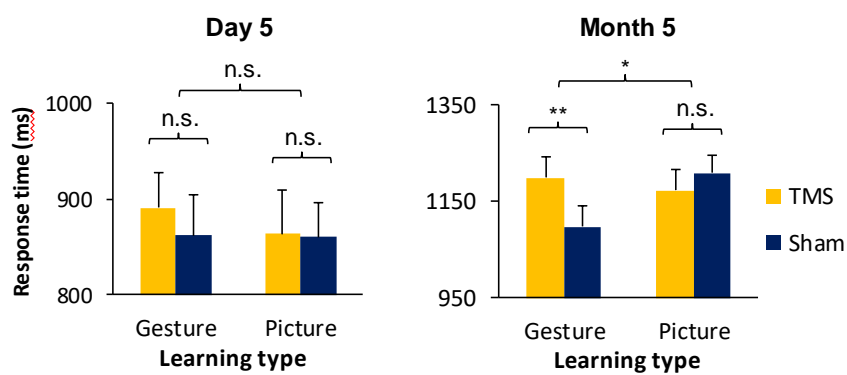

**Figure 3. Effects of bmSTS stimulation on speed of L2 translation by time point.** Effects of bmSTS stimulation on response times in the multiple choice task occurred 5 months following learning ($n = 22$ participants). Error bars represent one standard error of the mean. *$p < .05$, **$p < .01$.

**Role of L2 vocabulary concreteness**

Next, we tested our third hypothesis that the disruptive effects of bmSTS stimulation would occur independent of whether a word was classified as concrete or abstract (see also supplementary introduction). A four-way ANOVA on translation response times in the multiple choice task yielded a significant four-way learning condition × stimulation type × time point × vocabulary type interaction ($F_{1, 21} = 5.24$, $p = .033$, two-tailed, $\eta_p^2 = .20$) (see **Table S5**, supplementary material, for the full set of ANOVA results). Tukey's HSD post-hoc tests revealed that concrete nouns paired with gestures during learning were translated significantly more slowly during bmSTS stimulation compared to sham stimulation at day 5 ($p = .05$, Hedge's $g = .31$; **Fig. 4**). Contrary to our hypothesis, this comparison was not significant for abstract nouns at day 5. At month 5, however, TMS significantly slowed the translation of both L2 word types following gesture-enriched learning (concrete words: $p = .002$, Hedge's $g = .44$; abstract words: $p < .001$,

Hedge's *g* = .48). Response times under effective and sham stimulation did not significantly differ for words of either type that were learned in the picture enrichment condition at either time point. In sum, stimulation of the bmSTS modulated the translation of the concrete gesture-enriched L2 vocabulary at the earlier time point, and the translation of both concrete and abstract gesture-enriched vocabulary at the later time point.

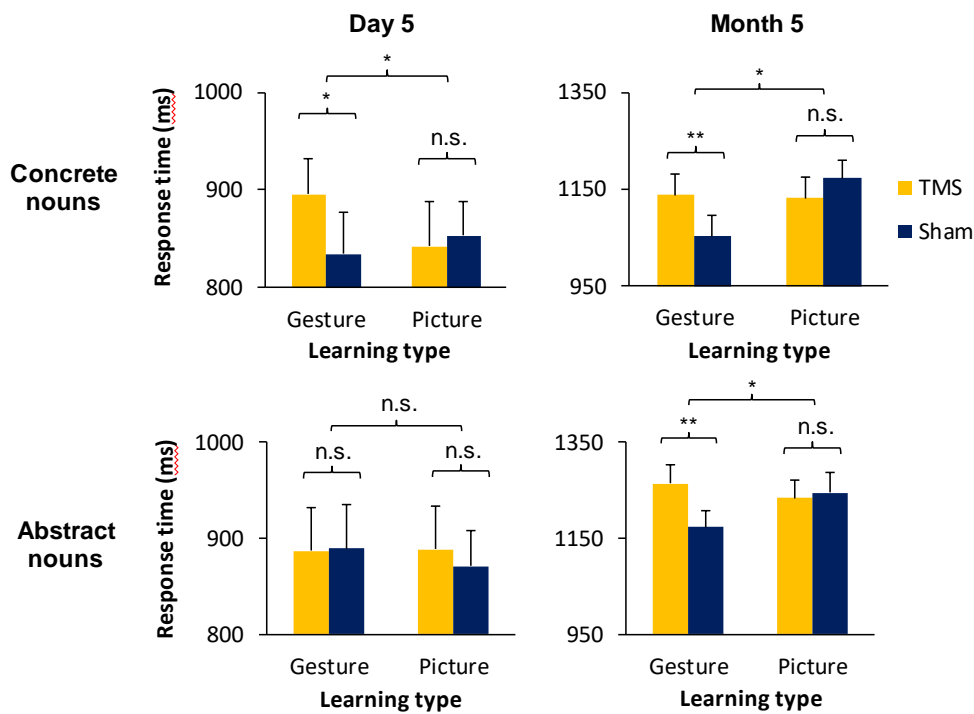

**Figure 4. Effects of bmSTS stimulation on speed of L2 translation by vocabulary type.** L2 vocabulary translation response times at the day 5 TMS session (left) and month 5 TMS session (right) by stimulation type, learning type, and vocabulary type (*n* = 22 participants). Compared to sham stimulation, stimulation of the bmSTS delayed response selection for concrete gesture-enriched nouns at day 5 and for both concrete and abstract gesture-enriched nouns at month 5. Error bars represent one standard error of the mean. *$p < .05$, **$p < .01$.

Testing of our three hypotheses using linear mixed effects modeling yielded qualitatively similar results as the ANOVA-based approach reported heretofore, with the exception of the four-way interaction, which was significant in the four-way ANOVA but not in the mixed effects model analysis. Please see the supplementary material (**Table S2**) for the full mixed effects model results.

**Gesture-enriched training facilitates long-term L2 translation accuracy**

Besides testing our main hypotheses related to effects of TMS on translation, the data also allowed us to test the reliability of the previous finding that benefits of gesture performance enrichment on L2 translation exceed those of picture viewing enrichment over the long-term (Mayer et al., 2015). To test this, we conducted a four-way ANOVA on translation accuracy scores in the multiple choice task (percent correct) with the factors learning condition, stimulation type, time point, and vocabulary type. The ANOVA revealed a significant learning condition × time point interaction ($F_{1, 21} = 6.86$, $p = .016$, two-tailed, $\eta_p^2 = .25$) (see **Table S6**, supplementary material, for the full set of ANOVA results). Tukey's HSD post-hoc tests revealed greater response accuracy following gesture-enriched learning compared to picture-enriched learning at month 5 ($p = .035$, Hedge's $g = .11$), which did not occur at day 5, suggesting that gesture-enrichment-based benefits on response accuracy emerged over a period of several months (**Fig. 5**). This finding is consistent with the previous report that gesture enrichment outperforms picture enrichment over longer timescales (Mayer et al., 2015).

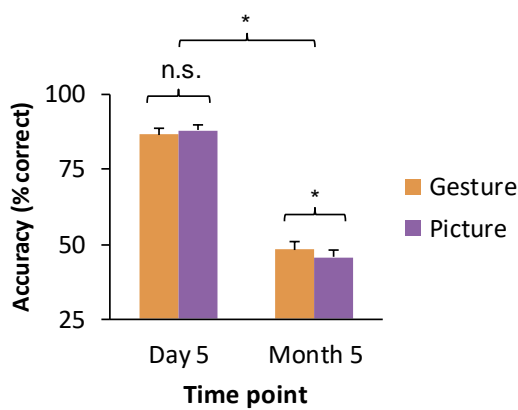

**Figure 5. Accuracy of L2 translation following learning.** Learning condition and time point variables in the multiple choice task significantly interacted: Participants translated gesture-enriched L2 words more accurately than picture-enriched L2 words at month 5 only ($n = 22$ participants).

We report here, for completeness, further tests and results related to multiple choice task accuracy. As expected, there were no significant effects of stimulation type on accuracy for either vocabulary type or learning condition at either time point (**Fig. 6**). However, there was an unexpected significant four-way learning condition × stimulation type × time point × vocabulary type interaction ($F_{1, 21} = 8.23$, $p = .009$, two-tailed, $\eta_p^2 = .28$) attributable to a significant difference in response accuracy between concrete and abstract gesture-enriched – but not picture-enriched – words at month 5 under sham stimulation ($p < .001$, Hedge's $g = .79$). Participants translated concrete words significantly more accurately overall in the multiple choice task than abstract words, a main effect of vocabulary type ($F_{1, 21} = 35.62$, $p < .001$, two-tailed, $\eta_p^2 = .63$). This effect was expected based on previous studies (Macedonia & Klimesch, 2014; Macedonia & Knösche, 2011),

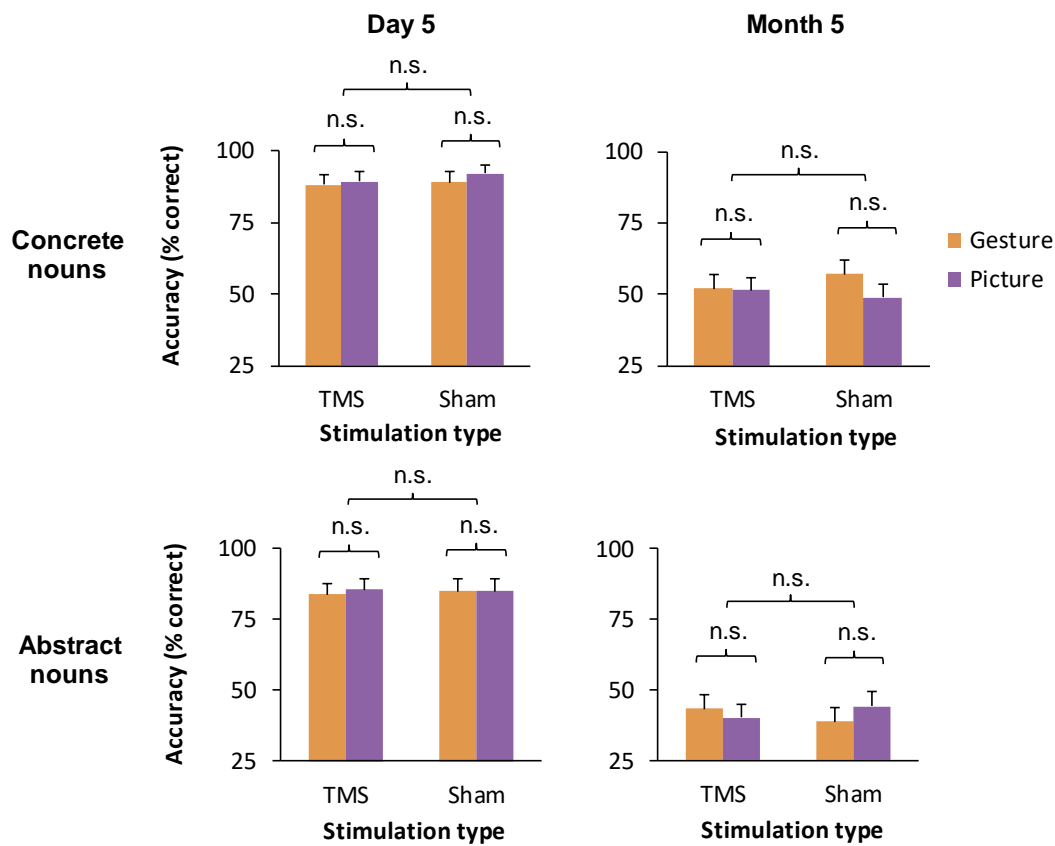

**Figure 6. Accuracy of L2 translation depending on learning condition, stimulation type, time point, and vocabulary type.** As expected, no significant effects of TMS on L2 translation accuracy in the multiple choice task were observed at either time point ($n = 22$ participants).

Participants were also significantly less accurate in the multiple choice task at month 5 compared to day 5, a main effect of time point ($F_{1, 21} = 124.77$, $p < .001$, two-tailed, $\eta_p^2 = .86$), suggesting that L2 memory representations decayed over time. To test whether L2 memory decayed less following gesture-enriched learning then picture-enriched learning, we computed changes in multiple choice task accuracy (percent correct) across the two time points (month 5 – day 5) in the absence of neurostimulation. A two-way ANOVA on changes in translation accuracy with the factors learning condition and vocabulary type yielded a significant learning condition × vocabulary type interaction ($F_{1, 21} = 13.84$, $p = .001$, two-tailed, $\eta_p^2 = .40$). Tukey's HSD post-hoc tests revealed a greater decrease in translation accuracy over the 5-month interval following picture-enriched learning compared to gesture-enriched learning for concrete words only ($p = .009$, Hedge's $g = .58$) (see **Table S7**, supplementary material, for the full set of ANOVA results). Thus, gesture-enriched representations of concrete L2 words decayed less than picture-enriched L2 representations (**Fig. S1a**, supplementary results). This result cannot be explained by a tradeoff in accuracy and response time, as the same two-way ANOVA on changes in response times (month 5 – day 5) with factors learning condition and vocabulary type yielded qualitatively similar results: Response times increased more over the 5-month interval following picture-enriched learning compared to gesture-enriched learning for both vocabulary types (**Fig. S1b**, supplementary results), a main effect of learning condition ($F_{1,21} = 11.05$, $p = .003$, two-tailed, $\eta_p^2 = .34$) (see **Table S8**, supplementary material, for the full set of ANOVA results). Thus, gesture-enriched learning benefitted translation response times more than picture-enriched learning over the long-term, further indicating the greater robustness of gesture-enriched L2 representations in memory over a long timescale (see supplementary results for more details).

Participants' scores on the paper-and-pencil vocabulary tests, which were completed on days 2, 3, and 4 of the L2 training period at 5 months post-training, converged on a similar pattern of results as the multiple choice decay analyses (for details, please see the supplementary results, **Fig. S2** and **Fig. S3**).

## Discussion

This study revealed causal links between responses in specialized sensory cortices and facilitative effects of sensorimotor-enriched learning. There were three main findings. First, behavioral benefits of gesture-enriched learning were caused in part by responses within a specialized visual brain area, the bmSTS; this area was causally engaged in the auditory translation of gesture-enriched but not picture-enriched L2 words. Second, bmSTS integrity supported the auditory translation of gesture-enriched words at 5 months post-learning even more than 5 days post-learning. Third, bmSTS integrity supported the translation of both concrete and abstract L2 words; stimulation effects were observed for concrete nouns at the earlier time point, and for both word types at the later time point. Taken together, these findings show that sensorimotor-enriched teaching constructs strong associations between auditory L2 words and their L1 translations by way of representations arising from specific visual cortices. Robust long-term memory representations established by sensorimotor-enriched teaching can therefore be supported by task-specific, specialized sensory brain responses.

The causal relation observed between bmSTS responses and L2 translation adjudicates between influential reactivation (Fuster, 2009; Nyberg et al., 2000; Wheeler et al., 2000) and predictive coding (Friston, 2012; von Kriegstein, 2012) theories of multisensory learning. The fact that brain responses in one sensory modality (e.g., visual) can improve task performance in another modality (e.g., auditory), depending on associations forged during learning, is expected based on predictive coding theories but not reactivation theories. Reactivation theories do not, in general, consider the idea that reactivated brain regions could serve some kind of role in perception. Given that knowledge of the gesture associated with an L2 word was not critical for achieving success in the multiple choice task, reactivation theories would presume that bmSTS integrity would not contribute to task performance and that inhibitory stimulation of the bmSTS would not interfere with completing the task. Though both gesture- and picture-enriched training involved complementary visual information, disruptive effects of bmSTS stimulation occurred only for the condition that contained stimulus information related to biological motion. Therefore, bmSTS engagement depended on sensorimotor experience. Based on predictive coding theories, we would expect motor and somatosensory stimulation to similarly disrupt the translation of gesture- but not picture-enriched words (for preliminary results, see Mathias et al., 2019), and expect LOC stimulation to disrupt the translation of picture- but not gesture-enriched words. Our "virtual lesion" TMS approach took advantage of the focal spatial resolution of TMS to transiently interfere with processing in a specific cortical target (Sack et al., 2007). If the current results were due to a whole-brain effect of TMS rather than one localized to the STS region, TMS would have lengthened response times not only for gesture-enriched words, but also picture-enriched words.

The performance of visually-modeled gestures yielded beneficial long-term effects on L2 translation accuracy and lessened long-term L2 decay compared to picture-enriched learning. Gesture enrichment facilitated learning, in part, by establishing representations of learned information within specific visual cortices. In our experiment, we characterize "learning by doing" as "sensorimotor-enriched learning" rather than "motor-enriched learning" because motor components of gesture-based enrichment can never be fully separated from associated sensory components. Even if learners performed self-created gestures without viewing a model, they would still receive visual feedback from their own and others' body movements, as well as other types of movement-associated sensory feedback. Learning by doing inevitably involves the integration of sensory and motor aspects of one's experience.

In order to recall the meaning of a newly-acquired L2 word, the brain may internally simulate sensory and motor processes that were involved in learning that word. This view is consistent with the notion that the presence of additional dimensions (e.g., visual) along which stimuli can be evaluated during recognition underlie learning-by-doing-based benefits (MacLeod, Gopie, Hourihan, Neary, & Ozubko, 2010). Mayer and colleagues (2015) found that viewing videos of gestures did not benefit post-training performance compared to auditory-only learning. However, behavioral outcomes following the video-viewing condition also correlated with decoded bmSTS responses measured using fMRI. This could indicate that the bmSTS is also engaged if learning involves viewing gestures and that other regions encode the visually-enriched vocabulary less efficiently, but that the bmSTS is unable to compensate for these deficiencies. Whether bmSTS stimulation would disrupt the translation of vocabulary learned by viewing gestures (and not performing them) is an open question. On the basis of previous fMRI results (Mayer et al., 2015), we reason that stimulation would disrupt performance.

A growing literature has reported positive effects of arousal-based interventions such as physical exercise (Hötting, Schickert, Kaiser, Röder, & Schmidt-Kassow, 2016), emotion regulation (Storbeck, & Maswood, 2016), and even music (Schellenberg, Nakata, Hunter, & Tamoto, 2007) on cognitive task performance. Though effective, these approaches do not encode associations between different components of the word acquisition experience in the same way as gesture-enriched learning, as gestures are intrinsically bound to specific stimulus information (Markant, Ruggeri, Gureckis, & Xu, 2016). If behavioral benefits of enrichment were due solely to increased arousal during gesture-enriched learning compared to picture-enriched learning, then stimulation of a specialized visual area would not have disrupted those benefits. Further, any potential differences between gesture- and picture-enriched learning in terms of arousal were not large enough to distinguish these conditions in terms of performance accuracy. Previously, the combination of a motor task with picture viewing (tracing an outline of presented pictures) during L2 learning benefitted learning outcomes less than simply viewing pictures without performing any movements (Mayer et al., 2015), and the performance of semantically-related gestures enhanced learning outcomes compared to the performance of meaningless gestures (Macedonia, Müller, & Friederici, 2011). These outcomes suggest that gesture enrichment benefits cannot be explained simply by the presence of movement during learning. The current results therefore steer away from more general explanations for beneficial effects of sensorimotor-enriched or multisensory-enriched learning such as increased arousal or attention. Hence, teaching strategies may be advanced by establishing links between new information and congruent sensorimotor and multisensory enrichment.

We conclude that sensorimotor-enriched training constructs stronger associations between auditory L2 words and their L1 translations than commonly-practiced sensory-only methods in adults. Beneficial behavioral effects of sensorimotor-enriched training are caused in part by responses within specialized sensory brain regions. Spoken language perception may therefore rely not only on auditory information stored in memory, but also on the sensorimotor context in which words are experienced. The causal relation observed between sensory brain responses and behavioral performance significantly advances our knowledge of neuroscientific mechanisms contributing to benefits of sensorimotor-enriched learning. The current findings also shed new light on the idea that sensorimotor-enriched teaching practices can be used to enhance learning outcomes by linking sensory brain functions with behavioral performance, and may have repercussions for the ways in which current classroom teaching practices are evaluated.

## Acknowledgements

This work was funded by German Research Foundation grant KR 3735/3-1, a Max Planck Research Group grant to K.v.K, and an Erasmus Mundus Postdoctoral Fellowship. B.M. is also supported by the European Research Council Consolidator Grant SENSOCOM 647051 to K.v.K. We thank Frieder Schillinger for comments on a previous version of the manuscript.

## Author Contributions

M.M., K.M.M., and K.v.K. conceived the study. B.M., G.H., K.v.K. developed the study design. B.M. and L.S. acquired and analyzed the data. B.M. and K.v.K. wrote the manuscript. L.S., G.H., M.M., and K.M.M. contributed to writing the manuscript.

## Open Practices Statement

The experiment reported in this article was not formally preregistered. The hypotheses were derived directly from the findings of a previous paper (Mayer et al., 2015). Neither the data nor the materials have been made available on a permanent third-party archive; requests for the data or materials can be sent via email to the lead author at brian.mathias@tu-dresden.de.

**Supplementary Material**

**Influences of the conceptual perceptibility of L2 word referents**

    **Introduction.** A potentially limiting factor in the success of sensorimotor-enriched approaches to L2 vocabulary learning may arise from the conceptual perceptibility of word referents. Conceptual perceptibility refers to the extent to which referents can be perceived by the body's sensory systems (e.g., tangibility; Hoffman, 2016). The referent of the concrete noun *ball*, for example, is highly tangible and can be iconically represented by using one's arms to throw an imaginary ball. Referents of other words, such as the abstract noun *mentality*, are less tangible and more difficult to convey using gestures or pictures. Despite differences in terms of intrinsic sensorimotor associations, the learning of both concrete and abstract vocabulary has been shown to benefit from gesture and picture forms of enrichment (Macedonia, 2014; Mayer, Yildiz, Macedonia, & von Kriegstein, 2015), suggesting that sensory brain regions should contribute to learning benefits for concrete and abstract words. Sensorimotor facilitation of the learning of abstract nouns may function by building associations between abstract concepts and perceptible sensory and motor events. The L2 translation of the word *innocence*, for example, is difficult to learn if paired simply with its native language translation. It becomes easier to learn if paired with the iconic gesture of shrugging of one's shoulders, even though *innocence* is not defined as shrugging (Macedonia & Knösche, 2011). Given that enriched learning may establish sensorimotor associations with both concrete and abstract words, the third aim of the current study was to address whether sensory brain regions differentially contribute to gesture-enriched learning benefits for these word types. Given that gesture-enriched learning previously benefitted concrete and abstract words similarly (Mayer et al., 2015), we hypothesized that specialized visual sensory responses would contribute to the translation of both word types. The concrete and abstract vocabulary used in the study is shown in **Table S1**.

    **Discussion of findings.** We found that bmSTS stimulation inhibited the translation of only concrete words at the earlier time point, and of both gesture-enriched concrete and abstract words at the later time point. We can only speculate on why this might be the case. Concrete concepts may map more easily onto gestures compared to abstract concepts. This may have resulted in learners' greater reliance on alternate learning strategies for translation of abstract L2 words at the earlier time point. Nevertheless, bmSTS stimulation inhibited the translation of both gesture-enriched concrete and abstract words at the later time point, consistent with the previous demonstration of greater long-term memory benefits for both word types following gesture enrichment compared to picture enrichment (Mayer et al., 2015). This result suggests that, if

alternate strategies were used for the abstract words, they were used only in the short-term, and learners relied over the long-term instead on sensorimotor representations.

**Table S1.** Vocabulary used in the experiment. 90 Vimmi and German words, and their English translations. Assignment of words to the gesture learning condition and the picture learning control condition was counterbalanced across participants, ensuring that each Vimmi word was represented equally in both learning conditions.

| Concrete nouns | | | Abstract nouns | | |
|---|---|---|---|---|---|
| **German** | **English** | **Vimmi** | **German** | **English** | **Vimmi** |
| Ampel | traffic light | gelori | Absage | Cancellation | munopa |
| Anhänger | trailer | afugi | Alternative | Alternative | mofibu |
| Balkon | balcony | usito | Anforderung | requirement | utike |
| Ball | ball | miruwe | Ankunft | Arrival | matilu |
| Bett | bed | suneri | Aufmerksamkeit | Attention | fradonu |
| Bildschirm | monitor | zelosi | Aufwand | Effort | muladi |
| Briefkasten | letter box | abota | Aussicht | View | gaboki |
| Decke | ceiling | siroba | Befehl | Command | magosa |
| Denkmal | memorial | frinupo | Besitz | Property | mesako |
| Eintrittskarte | entrance ticket | edafe | Bestimmung | Destination | wefino |
| Faden | thread | kanede | Bitte | Plea | pokute |
| Fahrrad | bicycle | sokitu | Disziplin | Discipline | motila |
| Fenster | window | uribo | Empfehlung | recommendation | giketa |
| Fernbedienung | remote control | wilbano | Gedanke | Thought | atesi |
| Flasche | bottle | aroka | Geduld | Patience | dotewa |
| Flugzeug | airplane | wobeki | Gleichgültigkeit | Indifference | frugazi |
| Gemälde | painting | bifalu | Information | Information | sapezo |
| Geschenk | present | zebalo | Korrektur | Correction | fapoge |
| Gitarre | guitar | masoti | Langeweile | Boredom | elebo |
| Handtasche | purse | diwume | Mentalität | Mentality | gasima |
| Kabel | cable | zutike | Methode | Method | efogi |
| Kamera | camera | lamube | Mut | Bravery | wirgonu |
| Kasse | till | asemo | Partnerschaft | Partnership | nabita |
| Katalog | catalog | gebamo | Rücksicht | consideration | ukowe |
| Kleidung | clothes | wiboda | Sensation | Sensation | boruda |
| Koffer | suitcase | mewima | Stil | Style | lifawo |
| Maschine | machine | nelosi | Talent | Talent | puneri |
| Maske | mask | epota | Tatsache | Fact | botufe |
| Papier | paper | serawo | Teilnahme | Participation | pamagu |
| Reifen | tire | wasute | Tendenz | Tendency | pefita |
| Ring | ring | guriwe | Theorie | Theory | sigule |
| Rucksack | backpack | lofisu | Therapie | Therapy | giwupo |
| Sammlung | collection | etuko | Tradition | Tradition | uladi |
| Schlüssel | key | abiru | Triumph | Triumph | gepesa |
| Schublade | drawer | lutepa | Übung | Exercise | fremeda |
| Sonnenbrille | sunglasses | woltume | Unschuld | Innocence | dafipo |

| Spiegel | mirror | dubeki | Veränderung | Change | zalefa |
|---|---|---|---|---|---|
| Straßenbahn | tram | umuda | Verständnis | Sympathy | gorefu |
| Tageszeitung | daily newspaper | gokasu | Vorgehen | Procedure | denalu |
| Telefon | telephone | esiwu | Vorwand | Excuse | Pirumo |
| Teller | plate | buliwa | Warnung | Warning | Gubame |
| Teppich | carpet | batewo | Wohlstand | Wealth | Bekoni |
| Verband | bandage | magedu | Wohltat | Benefaction | migedu |

**Analysis of long-term decay of L2 translation speed and accuracy**

To test whether memory decay over time was significantly less severe following gesture-enriched learning compared to picture-enriched learning, we computed changes in multiple choice task accuracy (percent correct) across the two time points (month 5 – day 5). Only sham condition accuracy was evaluated, in order to assess differences between gesture- and picture-enriched learning in the absence of neurostimulation. If gesture-enriched learning results in more robust L2 representations overall than picture-enriched learning, one might expect those representations to also decay less over time. A two-way ANOVA on changes in multiple choice response time across the two testing time points (month 5 – day 5) with the factors learning condition and vocabulary type revealed a main effect of learning condition ($F_{1,21}$ = 11.05, $p$ = .003, two-tailed, $\eta_p^2$ = .34) (see **Table S7** for the full set of ANOVA results).. A greater increase in response times occurred over the 5-month interval following picture-enriched learning compared to gesture-enriched learning, for both concrete nouns and abstract nouns (**Fig. S1a**). There were no other significant main effects or interactions. Thus, gesture-enriched learning benefitted translation response times more than picture-enriched learning over the long-term, indicating the greater robustness of gesture-enriched L2 representations in memory over a long timescale.

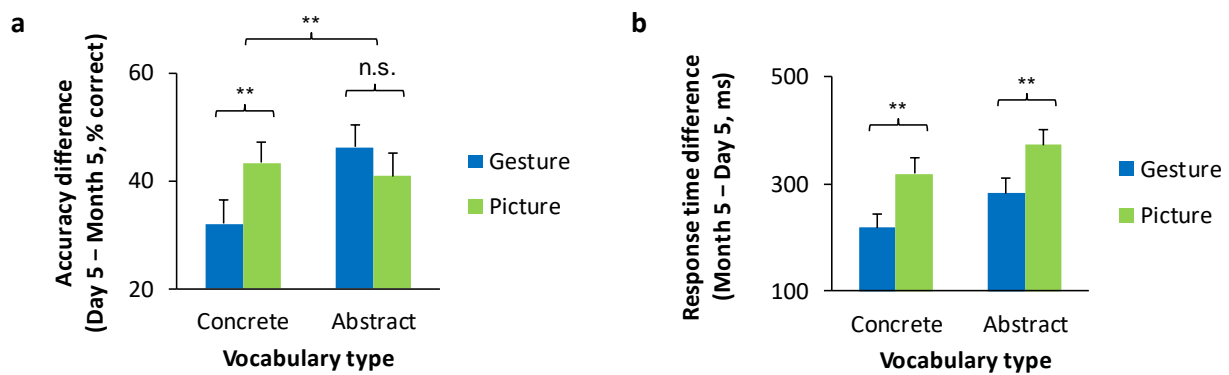

**Figure S1. Long-term change in accuracy and speed of L2 translation following learning. a,** Compared to traditional audiovisual (picture-enriched) learning, the performance

and viewing of iconic gestures during L2 vocabulary learning resulted in less decay of concrete L2 word knowledge over a 5-month period in the absence of neurostimulation compared to picture-enriched learning ($n$ = 22 participants). **b**, Gesture-enriched learning also resulted in less of an increase in translation response times at 5 months following training in the absence of neurostimulation ($n$ = 22 participants). Error bars represent one standard error of the mean. **$p < .01$.

This result cannot be explained by a tradeoff in accuracy and response time, as a two-way ANOVA on changes in response times (month 5 – day 5) with factors learning condition and vocabulary type the same two-way analysis of the response time variable (month 5 – day 5 difference) yielded qualitatively similar results: Response times increased more over the 5-month interval following picture-enriched learning compared to gesture-enriched learning for both vocabulary types (**Fig. S1b**, supplementary results), a main effect of learning condition ($F_{1,21}$ = 11.05, $p = .003$, two-tailed, $\eta_p^2 = .34$) (see **Table S8**, supplementary material, for the full set of ANOVA results). There were no other significant main effects or interactions. Thus, gesture-enriched learning benefitted translation response times more than picture-enriched learning over the long-term, further indicating the greater robustness of gesture-enriched L2 representations in memory over a long timescale.

**Analysis of response times in the exploratory recall task**

In the recall task, response time was defined as the time elapsed between the start of the auditory L2 word presentation and the participant's indication by button press (prior to the appearance of the four response options) that they knew the L1 translation of the presented L2 word. Participants indicated that they recalled the L1 translation prior to the appearance of the four response options during fewer than half of all trials across the two TMS sessions ($M$ = 41.7% of trials, $SE$ = 4.5%), leaving an insufficient number of trials for analysis of this exploratory task component. We nevertheless explored the data and analyzed the recall response times for correct response trials (in response to a reviewer's request). In order to evaluate recall response times for correct response trials, we analyzed trials in which participants first indicated by button press that they recalled the L1 translation and subsequently selected the correct translation from the list of response options presented on the screen.

A four-way ANOVA on recall response times for correct response trials with factors learning condition, stimulation type, time point, and vocabulary type yielded a significant main effect of time point, ($F_{1, 21}$ = 86.66, $p < .001$, two-tailed, $\eta_p^2 = .80$). Recall response times were significantly faster at day 5 than month 5. There was, however, no significant main effect of vocabulary type ($p$ = .96), which was one of the most robust effects throughout our other dependent measure, i.e., the multiple choice task reported in the main manuscript. Recall response times for concrete words ($M$ = 1527 ms, $SE$ = 26 ms) did not differ from response times for abstract words ($M$ = 1526 ms, $SE$ = 23 ms). The ANOVA yielded a significant learning condition × vocabulary type interaction ($F_{1, 21}$ = 6.52, $p$ = .019, two-tailed, $\eta_p^2$ = .24), and significant learning condition × time point × vocabulary type interaction ($F_{1, 21}$ = 4.38, $p$ = .049, two-tailed, $\eta_p^2$ = .17). However, Tukey's HSD post-hoc tests revealed no significant differences between concrete and abstract noun response times within any time point or learning condition. The predicted two-way interaction between learning condition and stimulation type variables was also not significant ($p$ = .49). Response times did not significantly differ between any conditions (TMS-Gesture: $M$ = 1506 ms, $SE$ = 33 ms; Sham-Gesture: $M$ = 1536 ms, $SE$ = 32 ms; TMS-Picture: $M$ = 1532 ms, $SE$ = 37 ms; Sham-Picture: $M$ = 1533 ms, $SE$ = 36 ms). There were no other significant main effects or interactions.

Given that not even the robust difference between concrete and abstract vocabulary types emerged in this analysis of recall response times, we assume that the low response rate yielded an insufficient number of trials for analysis of this task component. An alternative interpretation is that there was no effect of bmSTS stimulation on this specific vocabulary task.

**Analysis of TMS effects using linear mixed effects modeling**

To evaluate the robustness of the observed effects using an alternate analysis technique, we also tested our three hypotheses using a linear mixed effects modeling approach. Linear mixed effects models were generated in R version 1.2.1335 using the 'lme4' package (Bates, Maechler, Bolker, & Walker, 2015). To select the random effects structure, we performed backwards model selection, beginning with a random intercept by subject, a random intercept by auditory stimulus, a random slope by subject for each of the four independent factors (stimulation type, learning condition, time point, and vocabulary type), and a random slope by stimulus for the stimulation type and time point factors. We removed random effects terms that accounted for the least variance one by one until the fitted mixed model was no longer singular, i.e. until variances of one or more linear combinations of random effects were no longer (close to) zero. The final model included three random effects terms: a random intercept by subject, a random intercept by stimulus, and a random slope by subject for the time point factor.

Contrasts were coded using simple coding, i.e. ANOVA-style coding, such that the model coefficient represented the size of the contrast from a given predictor level to the (grand) mean (represented by the intercept). The dependent measure was response times in the multiple choice translation task. Significance testing was performed using Satterthwaite's method implemented in the 'lmerTest' package, with an alpha level of α = 0.05 (Kuznetsova, Brockhoff, & Christensen, 2017). Post-hoc Tukey tests were conducted using the 'emmeans' package (Lenth, Singmann, Love, Buerkner, & Herve, 2019). The full model results are shown in **Table S2**.

We first examined whether bmSTS stimulation modulated L2 translation. The model revealed a significant interaction of stimulation type and learning condition factors ($\beta$ = -15.43, $t$ = -3.34, $p$ < .001, 95% CI [-24.50 -6.37]), confirming our main hypothesis. Tukey's HSD post-hoc tests revealed that response times for words that had been learned with gesture enrichment – but not picture enrichment – were significantly delayed when TMS was applied to the bmSTS compared to sham stimulation ($\beta$ = -42.99, $p$ = .006).

We next examined whether bmSTS integrity supported the auditory translation of gesture-enriched words at the later time point (5 months post-learning) even more than the earlier time point (5 days following the start of learning). The model revealed a significant three-way interaction of stimulation type, learning condition, and time point variables ($\beta$ = -12.41, $t$ = -2.69, $p$ < .001, 95% CI [-21.48 -3.35]). Tukey's HSD post-hoc tests revealed a response benefit (faster responses) for gesture-enriched learning compared to picture-enriched learning under sham stimulation 5 months following learning ($\beta$ = -85, $p$ = .001). The application of TMS to bmSTS negated this benefit: Responses were significantly slower for the gesture condition at month 5 during TMS compared to sham stimulation ($\beta$ = -68, $p$ = .023).

Finally, we examined whether the disruptive effects of bmSTS stimulation would occur independent of the conceptual perceptibility of the L2 word referents (i.e., whether a word was concrete or abstract). The four-way stimulation type × learning condition × time point × vocabulary type interaction was not reliable in the fitted model ($\beta$ = 7.03, $t$ = 1.52, $p$ = .13, 95% CI [-2.02 16.08]), suggesting that effects of bmSTS stimulation did not significantly differ across vocabulary types. In sum, linear mixed modeling yielded the same results as the ANOVA-based approach reported in the main manuscript, with the exception of the four-way interaction, which was significant in the ANOVA but not in the mixed model analysis.

**Table S2.** Linear mixed effects regression testing the effects of stimulation type, learning condition, time point, and vocabulary type on response times in the multiple choice translation task.

| Fixed effects | | | | | | Random effects | | | |
|---|---|---|---|---|---|---|---|---|---|
| | Estimate | *SE* | *t* | *p* | CI | | | Variance | *SD* |
| Intercept | 1032 | 32.42 | 31.82 | <.001 | 965.66, 1098.10 | Participant | Intercept | 21325 | 146.03 |
| Stimulation | 6.07 | 4.62 | 1.31 | .19 | -3.00, 15.13 | | Time | 3674 | 60.61 |
| Learning condition | 3.23 | 4.63 | .70 | .49 | -5.85, 12.31 | Stimulus | Intercept | 5061 | 71.14 |
| Time point | 161.5 | 1.39 | 11.65 | <.001 | 132.69, 189.71 | | | | |
| Vocabulary | -32.79 | 6.99 | -4.69 | <.001 | -46.52, -18.83 | | | | |
| Stimulation × Learning | -15.43 | 4.62 | -3.34 | <.001 | -24.50, -6.37 | | | | |
| Stimulation × Time | -.009 | 4.63 | -.002 | .99 | -9.08, 9.07 | | | | |
| Learning × Time | 11.51 | 4.62 | 2.49 | .013 | 2.44, 20.57 | | | | |
| Stimulation × Vocabulary | -.69 | 4.61 | -.15 | .88 | -9.73, 8.36 | | | | |
| Learning × Vocabulary | 7.26 | 4.63 | 1.57 | .12 | -1.82, 16.35 | | | | |
| Time × Vocabulary | -29.12 | 4.70 | -6.20 | <.001 | -38.34, -19.91 | | | | |
| Stimulation × Learning × Time | -12.41 | 4.62 | -2.69 | .007 | -21.48, -3.35 | | | | |
| Stimulation × Learning × Vocabulary | -4.55 | 4.62 | -.98 | .33 | -13.60, 4.51 | | | | |

| | | | | | |
|---|---|---|---|---|---|
| Stimulation × Time × Vocabulary | -3.38 | 4.62 | -.73 | .46 | -12.42, 5.67 |
| Learning × Time × Vocabulary | 4.94 | 4.63 | 1.07 | .29 | -4.14, 14.01 |
| Stimulation × Learning × Time × Vocabulary | 7.03 | 4.62 | 1.52 | .13 | -2.02, 16.08 |

**Analysis of paper-and-pencil test data**

    **Methods.** On days 2, 3, and 4 of the L2 vocabulary learning, participants completed paper-and-pencil vocabulary tests prior to the training. Participants completed free recall, L1 translation, and L2 translation tests on each day. During the translation tests, participants received a list of either the 90 German words or the 90 Vimmi words and were asked to write the correct translation next to each word. During the free recall test, participants received a blank sheet of paper and were asked to write down as many German words, Vimmi words, or any combination of a Vimmi word with its German translation that occurred during the learning as they could remember. The free recall test was always administered before the translation tests, and the order of the two translation tests was counterbalanced across days and participants.

    Paper-and-pencil tests were independently scored for accuracy by two raters. L1 and L2 translation tests were scored in terms of the total number of correct translations recalled in each test (one point for each correct translation). A Vimmi word was considered correct if the two independent raters agreed that the word that was written down was valid for the sound pronounced in the audio file according to German sound-letter-mapping. A German word was considered correct if a participant wrote down the German word that was assigned to the Vimmi word during learning or if a participant wrote down a synonym of the German word, according to a standard German synonym database (http://www.duden.de). Free recall were scored in terms of the number of translations (German-Vimmi or Vimmi-German word pairs), German words that were missing corresponding Vimmi words, and Vimmi words that were missing corresponding German words. Three points were given for each correct translation (German-Vimmi or Vimmi-

German word pair). One point was given for each correctly-recalled German word that was missing a corresponding Vimmi translation and vice versa.

**Effects of enrichment on paper-and-pencil vocabulary test accuracy.**

***Translation tests.*** To analyze the translation tests, percentages of correctly translated words were averaged across the two tests (as in Mayer et al., 2015) and submitted to a three-way ANOVA with the factors learning condition (gesture, picture), testing time point (day 2, day 3, day 4, month 5), and vocabulary type (concrete, abstract). The ANOVA did not yield any interactions of the learning condition factor with other variables, suggesting similar effects of gesture- and picture-enriched learning on vocabulary test performance. There was a significant main effect of testing time point ($F_{3, 63} = 94.28$, $p < .001$, two-tailed, $\eta_p^2 = .82$). Tukey's HSD post-hoc tests revealed that overall test scores at each time point differed significantly from test scores at all other time points (all $p$s < .001, Hedge's $g$ range: .62-2.55; **Fig. S2a**). The ANOVA additionally yielded a significant main effect of vocabulary type ($F_{1, 21} = 135.17$, $p < .001$, two-tailed, $\eta_p^2 = .87$) and a significant time point × vocabulary type interaction ($F_{3, 63} = 5.78$, $p = .001$, two-tailed, $\eta_p^2 = .22$). Overall, test scores were significantly higher for concrete nouns compared to abstract nouns. There were no other significant main effects or interactions.

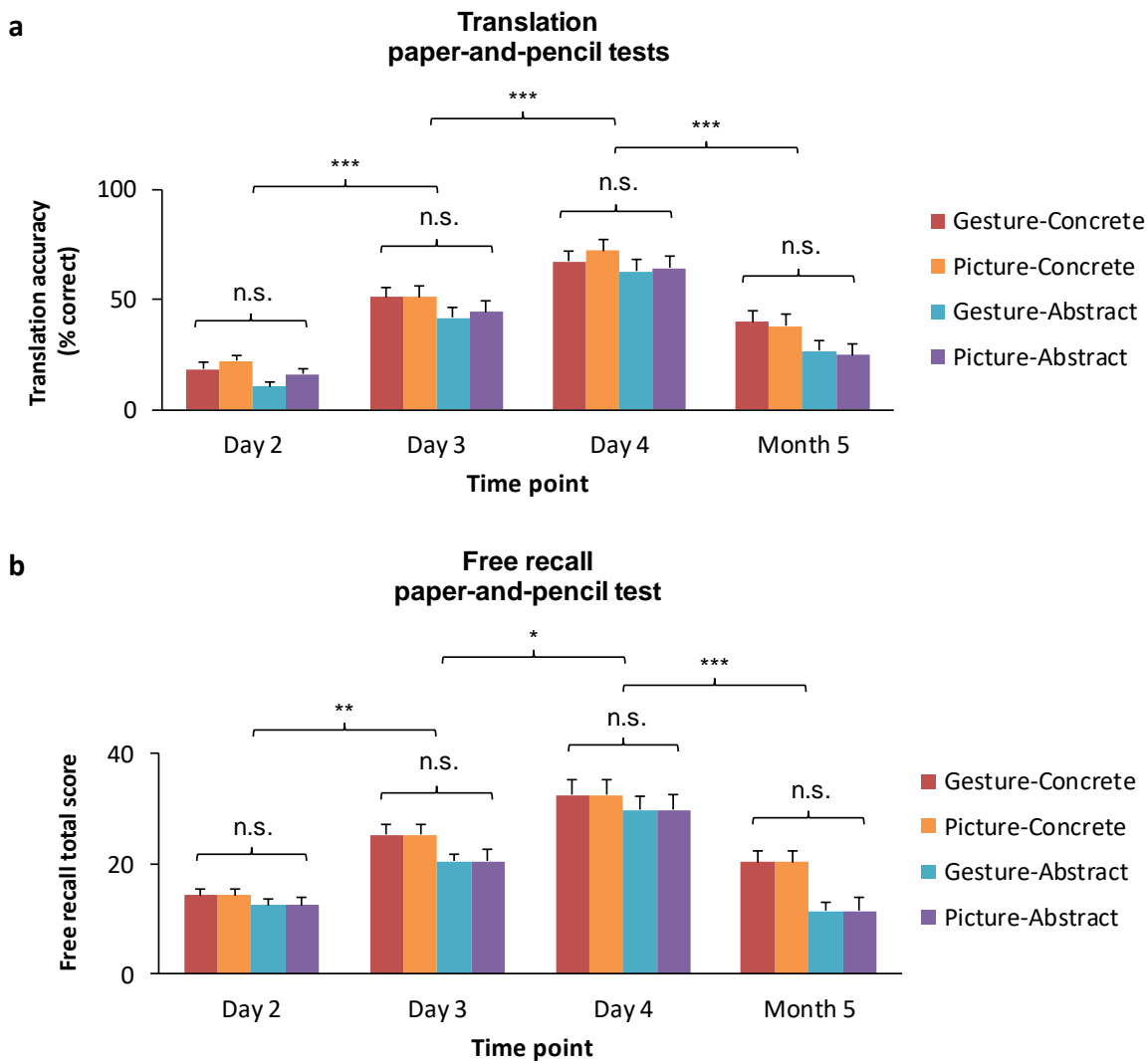

**Figure S2. Paper-and-pencil translation test scores. a,** Performance on paper-and-pencil translation tests significantly improved during days 2 to 4 of gesture- and picture-enriched training. Evidence of decay occurred 5 months following both gesture- and picture-enriched training (*n* = 22 participants). **b,** The same pattern of performance was observed for the free recall test (*n* = 22 participants). Error bars represent one standard error of the mean. \**p* < .05, \*\**p* < .01, \*\*\**p* < .001.

*Free recall test.* We next examined performance on the free recall paper-and-pencil test. Points for correctly recalled German words, Vimmi words, and German-Vimmi translations were summed for each participant, learning condition, testing time point, and vocabulary type (cf. Mayer et al., 2015). Free recall test scores by condition are shown in **Fig. S2b**. A three-way ANOVA on free recall scores with factors learning condition (gesture, picture), testing time point (day 2, day 3, day 4, month 5), and vocabulary type (concrete, abstract) did not yield any significant interactions of the learning condition factor with other variables besides a significant learning condition × vocabulary type interaction ($F_{1, 21} = 7.11$, $p = .014$, two-tailed, $\eta_p^2 = .25$). Tukey's HSD post-hoc tests revealed higher scores for concrete words compared to abstract words following gesture-enriched learning but not picture-enriched learning ($p < .001$, Hedge's $g$ = .40). There was a signficant main effect of vocabulary type ($F_{1, 21} = 11.14$, $p = .003$, two-tailed, $\eta_p^2 = .35$); scores were significantly higher for concrete words than abstract words. There was also a significant main effect of testing time point ($F_{3, 63} = 66.48$, $p < .001$, two-tailed, $\eta_p^2 = .76$). Tukey's HSD post-hoc tests revealed that overall test scores were significantly higher at day 3 compared to day 2 ($p = .0062$, Hedge's $g$ = 1.33), day 4 compared to day 3 ($p = .014$, Hedge's $g$ = .85), and at month 5 compared to day 4 ($p < .001$, Hedge's $g$ = 1.39). There was also a significant time point × vocabulary type interaction ($F_{3, 63} = 18.90$, $p < .001$, two-tailed, $\eta_p^2 = .47$). There were no other significant main effects or interactions.

**Gesture-enriched training reduces long-term decrease in L2 translation accuracy on paper-and-pencil vocabulary tests.** Finally, we tested whether gesture-enriched learning diminished long-term decreases in L2 translation accuracy over time compared to picture-enriched learning on the paper-and-pencil vocabulary tests.

*Translation tests.* In order to evaluate long-term changes in translation test accuracy, we computed the difference in mean performance on the translation tests (percent correct) at day 4 and month 5 for each participant, learning condition, and word type. A two-way ANOVA on difference scores (percent correct) with the factors learning condition (gesture, picture) and vocabulary type (concrete, abstract) yielded a significant main effect of learning condition ($F_{1, 21} = 5.84$, $p = .025$, two-tailed, $\eta_p^2 = .22$). Performance decreased significantly less for gesture-enriched vocabulary compared to picture-enriched vocabulary 5 months following training (**Fig. S3**). There was also a significant main effect of vocabulary type ($F_{1, 21} = 8.75$, $p = .007$, two-tailed, $\eta_p^2 = .29$). Performance decreased significantly less for concrete vocabulary compared to abstract vocabulary 5 months following training. The interaction between learning condition and vocabulary type variables was not significant.

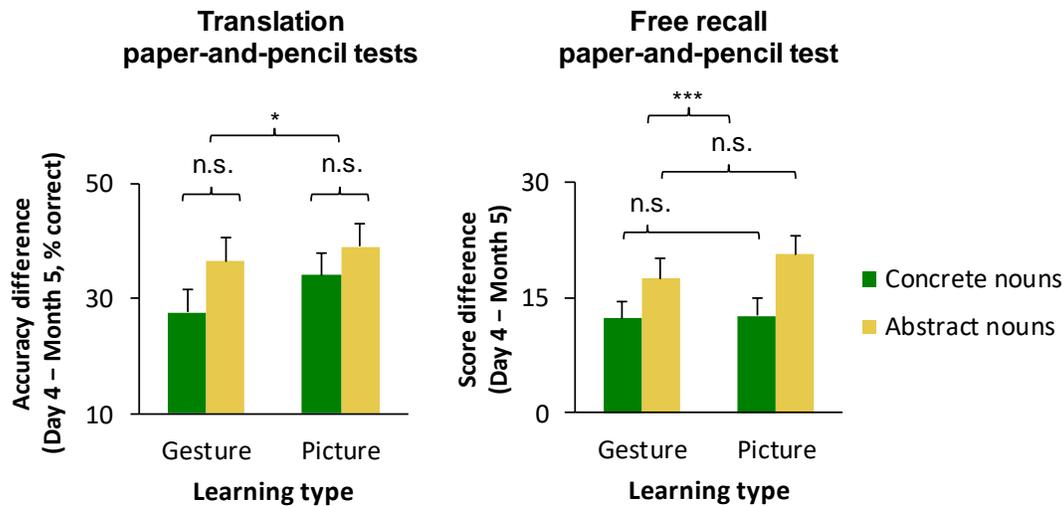

**Figure S3. Long-term decrease in translation accuracy on paper-and-pencil vocabulary tests.** Gesture-enriched L2 vocabulary learning resulted in less of a decrease in performance on paper-and-pencil translation tests 5 months following learning compared to picture-enriched learning. On the free recall paper-and-pencil test, participants demonstrated less long-term decay of concrete vocabulary compared to abstract vocabulary over a 5-month period ($n$ = 22 participants). Error bars represent one standard error of the mean. *$p$ < .05, **$p$ < .01, ***$p$ < .001.

*Free recall test.* In order to evaluate long-term changes in recall accuracy, we computed the difference in free recall paper-and-pencil test scores at day 4 and month 5 for each participant, learning condition, and word type. A two-way ANOVA on difference scores with the factors learning condition (gesture, picture) and vocabulary type (concrete, abstract) yielded only a significant main effect of vocabulary type ($F_{1, 21}$ = 18.43, $p$ < .001, two-tailed, $\eta_p^2$ = .47). Recall accuracy decreased significantly less for concrete vocabulary compared to abstract vocabulary 5 months following training (**Fig. S3**). The main effect of learning condition and interaction between learning condition and vocabulary type variables were not significant.

**Summary.** Taken together, the paper-and-pencil test scores revealed significant improvement for both gesture- and picture-enriched words from day 2 to day 3 and day 3 to day 4 of the L2 training period, as well as a significant decrease in performance 5 months post-learning compared to day 4. This pattern of performance was consistent across test types (translation and free recall tests). Analyses of L2 memory decay over a 5-month interval (day 4 scores – month 5 scores) revealed greater decay of memories for picture-enriched words compared to gesture-enriched words over a 5-month period on the translation tests. However, no difference between gesture- and picture-enriched words in terms of amount of decay was observed on the free recall tests. Instead, the free recall tests were sensitive to word type: A greater amount of decay was observed for abstract words compared to concrete words based on free recall scores.

## Analysis of variance (ANOVA) summary tables

In this section, we summarize using tables the main effects and interactions tested in all ANOVA analyses reported in the main manuscript. *p < .05, **p < .01, ***p < .001.

**Table S3.** Two-way ANOVA testing effects of stimulation type and learning condition on response times in the multiple choice translation task.

|  | df | F | p | $\eta_p^2$ |
|---|---|---|---|---|
| Intercept | 21 | 927.96 | <.001 | |
| Stimulation | 21 | 2.49 | .13 | .11 |
| Learning | 21 | .003 | .96 | <.001 |
| Stimulation × Learning | 21 | 11.82 | .002** | .36 |

**Table S4.** Three-way ANOVA testing effects of stimulation type, learning condition, and time point on response times in the multiple choice translation task.

|  | df | F | p | $\eta_p^2$ |
|---|---|---|---|---|
| Intercept | 21 | 1013.03 | <.001 | |
| Stimulation | 21 | 3.30 | .084 | .14 |
| Learning | 21 | 1.18 | .29 | .05 |
| Time | 21 | 18.65 | <.001*** | .84 |
| Stimulation × Learning | 21 | 106.95 | <.001*** | .47 |
| Stimulation × Time | 21 | 1.23 | .28 | .06 |
| Learning × Time | 21 | 6.42 | .019* | .23 |
| Stimulation × Learning × Time | 21 | 7.51 | .012* | .26 |

**Table S5.** Four-way ANOVA testing effects of stimulation type, learning condition, time point, and vocabulary type on response times in the multiple choice translation task.

|  | $df$ | $F$ | $p$ | $\eta_p^2$ |
|---|---|---|---|---|
| Intercept | 21 | 1055.04 | <.001 |  |
| Stimulation | 21 | 2.07 | .17 | .09 |
| Learning | 21 | .54 | .54 | .02 |
| Time | 21 | 131.51 | <.001*** | .86 |
| Vocabulary | 21 | 25.48 | <.001*** | .55 |
| Stimulation × Learning | 21 | 9.03 | .007** | .30 |
| Stimulation × Time | 21 | .014 | .91 | <.001 |
| Learning × Time | 21 | 8.29 | .009** | .28 |
| Stimulation × Vocabulary | 21 | .25 | .62 | .01 |
| Learning × Vocabulary | 21 | .03 | .86 | .001 |
| Time × Vocabulary | 21 | 3.30 | .083 | .14 |
| Stimulation × Learning × Time | 21 | 10.97 | .003** | .34 |
| Stimulation × Learning × Vocabulary | 21 | 2.95 | .10 | .12 |
| Stimulation × Time × Vocabulary | 21 | .44 | .52 | .02 |
| Learning × Time × Vocabulary | 21 | .11 | .74 | .005 |
| Stimulation × Learning × Time × Vocabulary | 21 | 5.24 | .033* | .20 |

**Table S6.** Four-way ANOVA testing effects of stimulation type, learning condition, time point, and vocabulary type on accuracy in the multiple choice translation task.

|  | $df$ | $F$ | $p$ | $\eta_p^2$ |
|---|---|---|---|---|
| Intercept | 21 | 361.33 | <.001 |  |

| | df | F | p | $\eta_p^2$ |
|---|---|---|---|---|
| Stimulation | 21 | 1.13 | .30 | .05 |
| Learning | 21 | .001 | .97 | <.001 |
| Time | 21 | 124.77 | <.001*** | .86 |
| Vocabulary | 21 | 35.62 | <.001*** | .63 |
| Stimulation × Learning | 21 | .035 | .85 | .002 |
| Stimulation × Time | 21 | .19 | .67 | .009 |
| Learning × Time | 21 | 6.86 | .016* | .25 |
| Stimulation × Vocabulary | 21 | 3.86 | .06 | .15 |
| Learning × Vocabulary | 21 | 1.25 | .28 | .06 |
| Time × Vocabulary | 21 | 3.93 | .90 | .16 |
| Stimulation × Learning × Time | 21 | .016 | .90 | <.001 |
| Stimulation × Learning × Vocabulary | 21 | 3.54 | .074 | .14 |
| Stimulation × Time × Vocabulary | 21 | .013 | .91 | <.001 |
| Learning × Time × Vocabulary | 21 | 2.60 | .12 | .11 |
| Stimulation × Learning × Time × Vocabulary | 21 | 8.23 | .009** | .28 |

**Table S7.** Two-way ANOVA testing effects of learning condition and vocabulary type on changes in multiple choice task accuracy across time points (month 5 – day 5).

| | df | F | p | $\eta_p^2$ |
|---|---|---|---|---|
| Intercept | 21 | 110.74 | <.001 | |
| Learning | 21 | 2.73 | .11 | .12 |
| Vocabulary | 21 | 3.41 | .079 | .14 |
| Learning × Vocabulary | 21 | 13.84 | .001*** | .40 |

**Table S8.** Two-way ANOVA testing effects of learning condition and vocabulary type on changes in multiple choice task response time across time points (month 5 – day 5).

|  | $df$ | $F$ | $p$ | $\eta_p^2$ |
|---|---|---|---|---|
| Intercept | 21 | 88.08 | <.001 |  |
| Learning | 21 | 11.05 | .003** | .34 |
| Vocabulary | 21 | 2.36 | .14 | .10 |
| Learning × Vocabulary | 21 | .017 | .90 | <.001 |